\documentclass[11pt]{article}
\usepackage{amsmath,amssymb}
\usepackage{bm}
\usepackage{graphicx}
\usepackage{hangcaption}

\title{Fast calculation of computer-generated-hologram on AMD HD5000 series GPU and OpenCL}
\author{Tomoyoshi Shimobaba$^*$, Tomoyoshi Ito$^*$, Nobuyuki Masuda$^*$, \\
Yasuyuki Ichihashi 
\thanks{Graduate School of Engineering, Chiba University, 1-33 Yayoi-cho, Inage-ku, Chiba 263-8522, Japan} 
and Naoki Takada
\thanks{Department of Imformation Media, Shohoku University, Atsugi, 243-8501 Japan}}

\date{\empty}
\begin{document}
\maketitle

\begin{abstract} 

In this paper, we report fast calculation of a computer-generated-hologram using a new architecture of the HD5000 series GPU (RV870) made by AMD and its new software development environment, OpenCL. 
Using a RV870 GPU and OpenCL, we can calculate $1,920 \times 1,024$ resolution of a CGH from a 3D object consisting of $1,024$ points in $30$ milli-seconds.
The calculation speed realizes a speed approximately two times faster than that of a GPU made by NVIDIA. 
\end{abstract}

\section{Introduction}
CGH (Computer-generated-hologram) has the ability to correctly  record and reconstruct a light wave for a 3D object.
Electroholography\cite{benton} using the CGH technique is attractive as a 3D display, because the CGH technique has remarkable features ; however, due to two significant problems, it is difficult to develop a practical 3D display system using electroholography.
One problem is the need for an SLM (spatial light modulator) that can display a CGH with large area and high resolution, because the resolution of a CGH is that of wavelength-order\cite{maeno, active,lee,takaki}. 
The other problem is the enormous computational time required for generating a CGH.
This paper focuses on this problem.

Assuming that a 3D object is composed of $N$ point light sources, the formula for computing a CGH is expressed as:
\begin{eqnarray}
I(x_h,~y_h)&=&\sum_j^N {A_j}{\rm cos}(\frac{2 \pi}{\lambda} (\frac{(p x_h-p x_j)^2+(p y_h-p y_j)^2}{2z_j}))\nonumber \\ 
&=&\sum_j^N {A_j}{\rm cos}(P_j ((x_h-x_j)^2+(y_h-y_j)^2)),
\label{eqn:cgh_basic}
\end{eqnarray}
where,  $I(x_h, y_h)$ is the light intensity on a CGH,  $(x_h,y_h)$ and $(x_j,y_j,z_j)$ are the coordinates for the CGH and a 3D object, $A_j$ is the light intensity of the 3D object, $\lambda$ is the wavelength of the reference light, and $P_j=\pi p^2 / (\lambda z_j)$, where $p$ is the sampling interval on the CGH plane.
Note that the coordinates $(x_j, y_j)$ and $(x_h, y_h)$ are normalized by $p$. 
The computational complexity of the above formula is O($N N_x N_y$), where $N_x$ and $N_y$ are the horizontal and vertical sampling numbers of the CGH.
This creates very enormous computational complexity.

To solve this problem, several software approaches have been proposed: for example, recurrence approaches\cite{yoshi, matsu, recurrence}, and the look-up table methods\cite{lut, image_hol, lutgpu}.
Another approach to dramatically reduce the calculation time is the hardware approach, such as FPGA (field-programmable gate array) and GPU (graphics processing unit).
We have designed and built special-purpose computers for holography using FPGA technology, called HORN (HOlographic ReconstructioN) \cite{ shimo_unit, horn5, horn6}.
The FPGA-based approaches showed excellent computational speed, however, the approach has the following restrictions: the high cost for developing the FPGA board, long development time and technical know-how required for the FPGA technology.

GPU-based approaches have already been applied to the optics field. 
Especially, CGH calculations \cite{masuda3, Ahrenberg,lutgpu, Onural} and reconstruction calculations in digital holography \cite{ gwo, shimo_dhm} are used to accelerate the calculation.
In 2007, NVIDIA released a new architecture of GPU and its software development environment, CUDA (Compute Unified Device Architecture).
Using CUDA allows us to program GPU easier than prior software developments, such as HLSL, Cg language and so forth.
Since the release, many papers using NVIDIA GPU and CUDA have been published in optics.  

On the other hand, more recently in December 2009, a new GPU of the HD5000 series (RV870) made by AMD was released.
The RV870 GPU has new architecture and its software environment, OpenCL (Open Computing Language).
The architecture of the RV870 GPU is different from that of the NVIDIA GPU. 
The RV870 GPU has huge potential for fast calculation because one GPU chip has over 1,000 floating-point number processors, while one NVIDIA GPU chip has about 200 floating-point number processors.
However, fast CGH calculation using the RV870 GPU has not been reported so far.

In this paper, we report fast CGH calculation using RV870 GPU and OpenCL. 
Using these, we can calculate $1,920 \times 1,024$ resolution of a CGH from a 3D object consisting of $1,024$ points in $30$ milli-seconds.
To the best of our knowledge, this article is the first report of using the RV870 GPU and OpenCL in optics.
In addition, we compare the calculation performance between the RV870 GPU and the GPU made by NVIDIA.  

In Section 2, we describe a fast CGH calculation on AMD RV870 and OpenCL.
In Section 3, we show and compare the performance between the RV870 GPU and the GPU made by NVIDIA.
In Section 4, we conclude this work.

\section{Fast calculation of computer-generated-hologram on AMD RV870 and OpenCL}

\begin{figure}[htb]
\centerline{
\includegraphics[width=12cm]{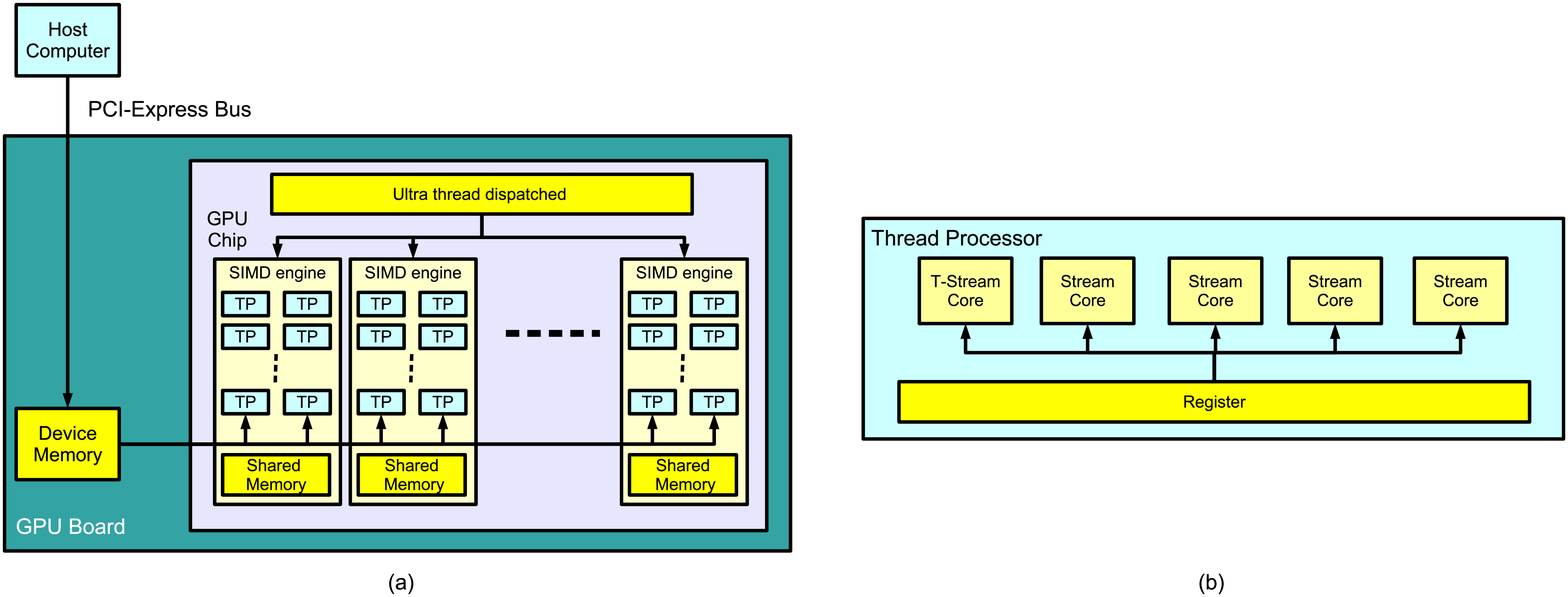}}
 \caption{Architecture of RV870 GPU. (a) Outline of RV870 GPU chip (b) Thread processor.}
\label{fig:amd}
\end{figure}

The architecture of RV870 GPU is shown in Fig.\ref{fig:amd}.
The top level of the GPU consists of many SIMD (Single Instruction Multiple Data) engines.
The SIMD engine has 16 thread processors (TP) and a shared memory, which is small and high-speed.

In addition, the thread processor has four stream cores and one T-stream core.
The stream core is a simple floating-point-number operation unit. 
And, the T-stream core also has a floating-point-number operation unit and special function unit.
The special function unit can calculate special functions at high speed, such as trigonometric function, logarithm function and so on.
The stream cores in the same SIMD engine operate by the same instructions; therefore, the SIMD engine is similar to a SIMD  processor.

Calculation on the GPU using OpenCL is executed using the following steps: 
(1) We initialize a GPU using OpenCL API (Application Program Interface) functions. 
(2) We allocate the required amount of memory on a device memory in Fig.\ref{fig:amd}. The device memory is large amount, but large latency access of memory. 
(3) We send an input data to the device memory. 
(4) We send a kernel function from the host computer to the GPU. The kernel function is compiled to native code of GPU using the OpenCL compiler. The GPU executes the kernel function.
(5) We receive a calculated result from the device memory.
(6) We release the device memory and GPU resources.

\begin{figure}[htb]
\centerline{
\includegraphics[width=15cm]{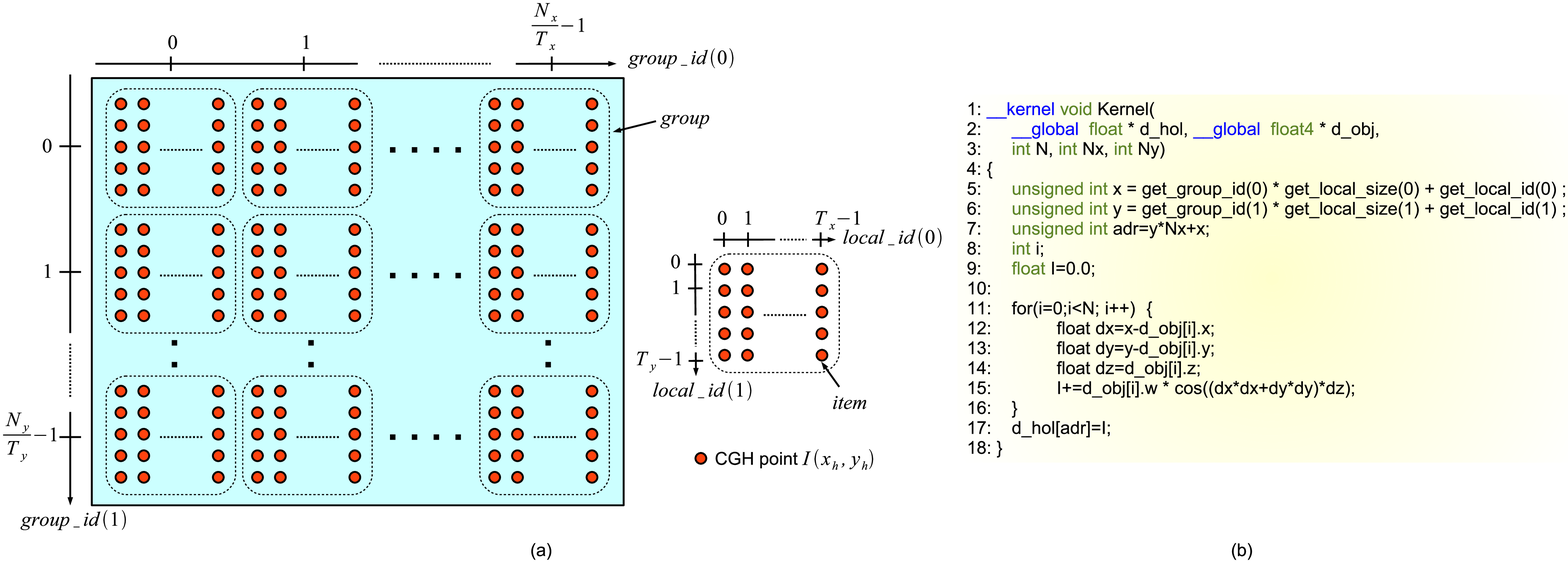}}
 \caption{(a) Outline of CGH calculation on RV870 GPU and OpenCL. (b) Kernel for the CGH calculation using OpenCL without optimization.}
\label{fig:block_list}
\end{figure}

Figure \ref{fig:block_list}(a) shows the outline of the CGH calculation on the RV870 GPU with OpenCL.
When calculating a CGH with the resolution of $N_x \times N_y$, we need to divide the CGH area into $group$s with the size of $T_x \times T_y$.
Therefore, the number of $group$s is $N_x/T_x \times N_y/T_y$.
In addition, each $group$ has $T_x \times T_y$ $item$s
(In the CUDA, $group$ and $item$ are equivalent to $block$ and $thread$, respectively).
Each $group$ is allocated to SIMD engines and each $item$ simultaneously calculate Eq.(\ref{eqn:cgh_basic}) by each stream core on an SIMD engine.

In Fig.\ref{fig:block_list}(b), we show the kernel source code of the CGH calculation on the RV870 GPU with OpenCL.
The source code is not optimized because we understand it easily.
The optimization is shown in the next subsection.

Each $group$ and $item$ have the indices, group\_id and local\_id.
The OpenCL functions, get\_group\_id(0) and get\_group\_id(1), give us the horizontal and vertical indices of group\_ids respectively. 
The OpenCL functions, get\_local\_id(0) and get\_local\_id(1), also give us the horizontal and vertical indices of local\_ids respectively. 

The arguments of the kernel function are a CGH data ($d\_hol$), an object data ($d\_obj$), the number of object points ($N$) and the CGH size ($N_x, N_y$).
An object data ($d\_obj$) consists of the coordinates and the intensity as four float data ($float4$).
In lines 5, 6 and 7 of the Fig.\ref{fig:block_list}(b), the variables $x$ and $y$ calculate the coordinates ($x_h, y_h)$ on the CGH plane and $adr$ calculates the address of the device memory for storing the calculation result $I(x_h,y_h)$.
In lines 11 to 16, a CGH point $I(x_h,y_h)$ can be calculated by iterating for $N$.
Although seeming to execute only one kernel, in fact, each stream core corresponding to $local\_id$ and $global\_id$ can perform the kernel in parallel.

When calculating a CGH with $1,920 \times 1,024$ from a 3D object composed of $1,024$ points, the kernel with $T_x \times T_y=16 \times 16$ took about $\rm 215 ms$.
The calculation speed of the kernel is slow.

\subsection{Optimization}
\label{sec:opt}

\begin{figure}[htb]
\centerline{
\includegraphics[width=13cm]{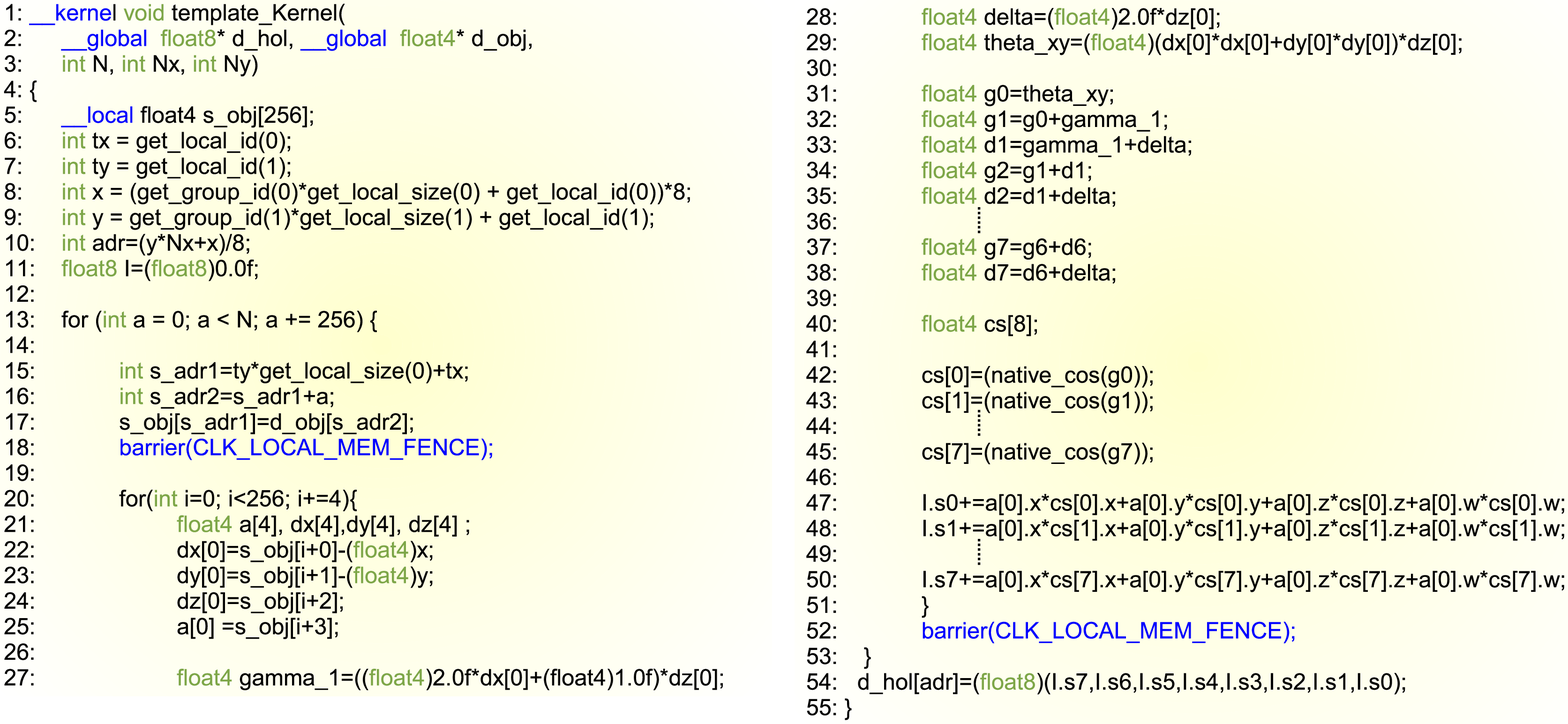}}
 \caption{Kernel for the CGH calculation using OpenCL with optimization.}
\label{fig:list2}
\end{figure}

The previous source code is not optimized.
In this section, we optimize the previous source code to obtain more acceleration speed.
Figure \ref{fig:list2} shows the optimized kernel function from Fig.\ref{fig:block_list}(b).

For more acceleration, we applied the following optimization techniques to the previous kernel: (1) Recurrence algorithm (2) Shared memory (3)  Loop unrolling   (4) Vectorization (5) Native instruction.

We proposed a fast CGH computation method using two recurrence formulas \cite{recurrence, shimo_unit, horn5,horn6}. 
Our recurrence algorithm can compute the phase component of the cosine function in Eq.(\ref{eqn:cgh_basic}) by two recurrence formulas.
The recurrence algorithm is as follows:

\begin{equation}
I(x_h+n, y_h)=\sum_j^N {A_j}{\rm cos}(\Gamma_n) ~~~ \Gamma_n=\Gamma_{n-1}+\delta_{n-1} ~~~  \delta_{n}=\delta_{n-1}+\Delta
\label{eqn:cgh_rec}
\end{equation}

Here, we define $\Gamma_0=P_j((x_h-x_j)^2+(y_h-y_j)^2)$, $\delta_0 = P_j (2 (x_h-x_j) + 1)$, $\Delta=2P_j$.
Eventually, we can compute the phase $\Gamma_n$ at the next coordinate by the two recurrence formulas.
For more details, see Ref.\cite{recurrence}

In lines 15 to 18, we copy the object data from the device memory ($d\_obj$) to a shared memory ($s\_obj$). 
The shared memory can store 256 object points at a time because the shared memory is small and high-speed.
Therefore, in the 13, we must iterate $N/256$ times.
Note that $barrier(CLK\_LOCAL\_MEM\_FENCE)$ means a barrier synchronization in line 18. 
It is equivalent to the $syncthreads$ function in the CUDA.

Loop unrolling is a well-known technique for optimizing a kernel function.
It can be realized by reducing the number of iterations and replicating the body of the loop.
Benefits of the loop unrolling are the capable to decrease the loop frequency, branch instructions and conditional instructions.
In the optimized kernel, we applied the loop unrolling to the loop of object points.
In lines 20 to 51 in Fig.\ref{fig:list2},  we can perform four object points per one iteration of the loop. 
In addition, we vectorize the operations in the loop using the $float4$ type, in order to handle four object points at a time.
For example, in line 22, we can calculate the four subtractions simultaneously. 
In the same way, the kernel can handle eight CGH points using the $float8$ type at same time in lines 40 to 50.

In lines 42 to 45, we used native cosine functions, instead of the normal cosine function shown in Fig.\ref{fig:block_list}(b).
The native cosine function can compute the fast cosine function using the hardware.

\subsection{Results}

Table \ref{tbl:time} shows a comparison of the calculation times for a CPU alone, NVIDIA GPU and an AMD RV870 GPU. 
The size of the CGH is $1,920 \times 1,024$.
The specifications of the personal computer are as follows: Intel Core 2 Quad Q6600 (We used one core for the calculation), 2 GB of memory, Microsoft Windows XP SP3. 
We used a GeForce GTX260 as the NVIDIA GPU board and its software development environment of CUDA version 2.3, and a RADEON HD5850 as the AMD GPU board and its software development environment of StreamSDK version2.0. 
The RADEON HD5850 GPU has 1,440 stream cores (namely, 18 SIMD engines) with the clock frequency of 725MHz.

We can see that the optimization method for the AMD GPU described in Section \ref{sec:opt} can perform more than ten times faster than that without the optimization.
In the calculation times for the NVIDIA GPU in the table, we optimized the kernel for the NVIDIA GPU using the same method as described in Section \ref{sec:opt}: namely, recurrence algorithm, shared memory, loop unrolling, vectorization, native instruction.
And, in the calculation times for the CPU alone in the table, we used Eq.(\ref{eqn:cgh_rec}) for the CGH calculation.
All calculation times using AMD and NVIDIA are superior to those using the CPU alone.
In addition, the AMD GPU can calculate a CGH approximately two times faster than the NVIDIA GPU.

\begin{table}[htb]
\caption{Comparison of calculation times on CPU alone, NVIDIA GPU and AMD GPU.}
\begin{center}
\scalebox{0.7}{
\begin{tabular}{|c|c|c|c|c|}
\hline
\multicolumn{ 1}{|c|}{Number of object points} & \multicolumn{ 4}{c|}{Time(ms)} \\ \cline{ 2- 5}
\multicolumn{ 1}{|c|}{} & \multicolumn{ 1}{c|}{CPU} & \multicolumn{ 1}{c|}{NVIDIA GPU} & \multicolumn{ 2}{c|}{AMD GPU} \\ \cline{ 4- 5}
\multicolumn{ 1}{|c|}{} & \multicolumn{ 1}{c|}{} & \multicolumn{ 1}{c|}{} & \multicolumn{1}{c|}{Without optimization} & \multicolumn{1}{c|}{With optimization} \\ \hline
512 & $30 \times 10^3$ & 33 & 215 & 19 \\ \hline
1024 & $59 \times 10^3$ & 59 & 422 & 31 \\ \hline
1536 & $88 \times 10^3$ & 85 & 630 & 44 \\ \hline
2048 & $115 \times 10^3$ & 112 & 838 & 56 \\ \hline
2560 & $146 \times 10^3$ & 139 & 1045 & 68 \\ \hline
3072 & $174 \times 10^3$ & 165 & 1252 & 80 \\ \hline
3584 & $204 \times 10^3$ & 192 & 1464 & 93 \\ \hline
\end{tabular}
}
\end{center}
\label{tbl:time}
\end{table}

\section{Conclusion}

In this paper, we described a fast CGH calculation using an AMD RV870 GPU with new architecture and its new software development environment, OpenCL. 
Many fast CGH calculation methods using a NVIDIA GPU and the CUDA have already been reported in optics field; however, a study using the RV870 GPU has not been reported so far.
To the best of our knowledge, this article is the first report of using the RV870 GPU and OpenCL in optics.
Using the RV870 GPU and OpenCL, we can calculate $1,920 \times 1,024$ resolution of a CGH from a 3D object consisting of $1,024$ points in about 30 ms.
The calculation speed can realize approximately two times faster than the NVIDIA GPU. 

This research was partially supported by the Ministry of Internal Affairs and Communications, Strategic Information and Communications R\&D Promotion Programme (SCOPE), 2009, and Japan Society for the Promotion of Science (JSPS), Grant-in-Aid for Scientific Research
(C) (21500094).


\begin{thebibliography}{99}

\bibitem{benton}
S. A. Benton, ``Experiments in holographic video imaging,'' Proc.SPIE {\bf IS8}, 247--267 (1991).

\bibitem{maeno}
K. Maeno, N. Fukaya, O. Nishikawa, K. Sato and T. Honda, ``ELECTRO-HOLOGRAPHIC display using 15MEGA pixels LCD,'' Proc.SPIE {\bf 2652}, 15--13 (1996).

\bibitem{active} 
C. W. Slinger et al., ``Recent Developments in Computer-Generated Holography: Toward a Practical Electroholography System for Interactive 3D Visualization,'' Proc SPIE., {\bf 5290}, 27--41 (2004).


\bibitem{lee}
J. Hahn, H. Kim, Y. Lim, G. Park and B. Lee, ``Wide viewing angle dynamic holographic stereogram with a curved array of spatial light modulators,'' Opt. Express, {\bf 16}, 12372--12386 (2008). 

\bibitem{takaki}
Y. Takaki and N. Okada, ``Hologram generation by horizontal scanning of a high-speed spatial light modulator,'' Appl. Opt., {\bf 48}, 3256--3261 (2009). 



\bibitem{yoshi} 
H.Yoshikawa, ``Fast Computation of Fresnel Holograms Employing Difference,'' Opt.Rev., {\bf 8,} 331 (2000).

\bibitem{matsu} 
K. Matsushima and M. Takai, ``Recurrence Formulas for Fast Creation of Synthetic Three-Dimensional Holograms,'' Appl. Opt., {\bf 39,} 6587 (2000).

\bibitem{recurrence} 
T. Shimobaba and T. Ito, ``Special-purpose computer for holography HORN-4 with recurrence algorithm,'' Comp. Phys. Commun., {\bf 148}, 160--170 (2002).

\bibitem{lut} 
M. Lucente, ``Interactive Computation of holograms using a Look-up Table,''  J. Electron. Imaging, {\bf 2}, 28--34 (1993).

\bibitem{image_hol} 
H. Yoshikawa, T. Yamaguchi, and R. Kitayama, ``Real-Time Generation of Full color Image Hologram with Compact Distance Look-up Table,''  OSA Topical Meeting on Digital Holography and Three-Dimensional Imaging 2009, DWC4 (2009). 

\bibitem{lutgpu} 
Y. Pan, X. Xu, S. Solanki, X. Liang, R. Bin A. Tanjung, C. Tan, and T. C. Chong,''  Fast CGH computation using S-LUT on GPU,''  Opt. Express, {\bf 17}, 18543--18555 (2009).

\bibitem{shimo_unit}
T. Shimobaba, A. Shiraki, N. Masuda, and T. Ito, ``Electroholographic display unit for three-dimensional display by use of special-purpose computational chip for holography and reflective LCD panel,''  Opt. Express,  {\bf  13}, 4196--4201 (2005).

\bibitem{horn5}
T. Ito, et.al., ``Special-purpose computer HORN-5 for a real-time electroholography,''  Opt. Express, {\bf 13}, 1923--1932 (2005).

\bibitem{horn6}
Y. Ichihashi, H. Nakayama, T. Ito, N. Masuda, T. Shimobaba, A. Shiraki and T. Sugie ``HORN-6 special-purpose clustered computing system for electroholography,''  Opt. Express, {\bf 17}, 13895--13903 (2009).

\bibitem{masuda3}
N. Masuda, T. Ito, T. Tanaka, A. Shiraki and T. Sugie, ``Computer generated holography using a graphics processing unit,''  Opt. Express, {\bf 14,} 2, pp.587--592 (2008).

\bibitem{Ahrenberg}
L. Ahrenberg, P. Benzie, M. Magnor, J. Watson, ``Computer generated holography using parallel commodity graphics hardware,'' Opt. Express, {\bf 14,} 17, pp.7636--7641 (2006).

\bibitem{Onural}
H. Kang, F. Yaras, and L. Onural, ``Graphics processing unit accelerated computation of digital holograms,'' Appl. Opt., {\bf 48}, H137--H143 (2009). 

\bibitem{gwo}
T. Shimobaba, et.al., ``Numerical calculation library for diffraction integrals using the graphic processing unit : the GWO library,'' J. Opt. A: Pure Appl. Opt., {\bf 10}, 075308 (5pp) (2008).

\bibitem{shimo_dhm}
T. Shimobaba, Y. Sato, J. Miura, M. Takenouchi and T. Ito, ``Real-time digital holographic microscopy using the graphic processing unit,''  Opt. Express,  {\bf  16}, 11776--11781 (2008)

\end{thebibliography}
\end{document}